# Determination of Interface Atomic Structure and Its Impact On Spin Transport Using Z-contrast Microscopy and Density-Functional Theory

Thomas J. Zega, Aubrey T. Hanbicki, Steven C. Erwin, Igor Žutić*, George Kioseoglou, Connie H. Li, Berend T. Jonker, and Rhonda M. Stroud

Naval Research Laboratory, Washington, D.C. 20375

*Current Address:
Department of Physics, State University of New York at Buffalo, Buffalo, NY 14260

**ABSTRACT**

We combine Z-contrast scanning transmission electron microscopy with density-functional-theory calculations to determine the atomic structure of the Fe/AlGaAs interface in spin-polarized light-emitting diodes. A 44% increase in spin-injection efficiency occurs after a low-temperature anneal, which produces an ordered, coherent interface consisting of a single atomic plane of alternating Fe and As atoms. First-principles transport calculations indicate that the increase in spin-injection efficiency is due to the abruptness and coherency of the annealed interface.

Obtaining an atomistic understanding of the effects of buried interfaces on electronic and magnetic properties is a long-standing problem in device physics. In general, the atomic structure and composition of interfaces are not known to the level of precision needed to perform accurate first-principles calculations of their electronic structure. Recent advances in high-angle annular-dark-field (HAADF) microscopy, the highest resolution, chemically sensitive transmission-electron-microscope (TEM) technique currently available [1-4], make it a very promising approach for addressing the buried-interface problem.

In principle, HAADF images allow direct quantitative determination of the atomic structure. However, direct interpretation is not always possible in the case of interfaces for which there may be abrupt changes in thickness, mixing of more than two elements on individual atomic columns, or when the probe-tail effects contribute to the signal on adjacent columns. Simulated HAADF images are essential for overcoming these difficulties because they allow quantitative comparison with experimental images for different imaging conditions. Plausible interface structures can be suggested by density-functional-theory (DFT) calculations, and thus, combined HAADF-DFT studies of buried interfaces are an important advance for relating interface structure to device properties.

Here we combine DFT with experimental- and simulated-HAADF imaging to determine the atomic structure of the interface between Fe and AlGaAs in light-emitting diodes (LEDs) used to investigate the transport of spin-polarized electrons [5-10]. Understanding spin transport is fundamental to spin-based electronics (spintronics), a new paradigm for semiconductor electronics in which electron spin, rather than charge, is utilized to carry and store information [11-13]. Calculations [14] and experiments [15] indicate that spin transport can be strongly influenced by the nature of the interface between the contact and the semiconductor.

Spin-polarized LED heterostructures were grown by molecular beam epitaxy and processed into surface emitting LEDs. Electrons were electrically injected from an Fe(001) film into an AlGaAs/GaAs(001) quantum well (QW), and the circular polarization of the emitted light provided a lower bound for the electron-spin polarization (SP) in the QW. No corrections were made for spin-lifetime effects. Further details of the growth, optical, and transport measurements may be found elsewhere [7, 16, 17]. Phase- and Z-contrast images were acquired along the [-110] zone axis with 200 keV JEOL 2010F and 2200FS TEMs ($C_s$ = 0.5 mm). HAADF imaging was performed in scanning mode using a sub-0.2-nm probe and 80- to 175-mrad collection semiangles. Experimental images were Fourier-filtered to improve the signal-to-noise ratio and compared to those simulated from four candidate interface models. We verified that the filtering did not introduce spurious periodicities to the image by comparing raw, Fourier-filtered, and low-pass filtered images, and used columns of adjacent Ga and As atoms in the bulk GaAs as an internal reference for the HAADF image contrast. Three of the interface models (abrupt, partially intermixed, and fully intermixed) were previously proposed and studied theoretically in Ref [18]. The fourth model contains several monolayers of $Fe_3GaAs$ (a known stable alloy in the Fe-Ga-As phase diagram) sandwiched as an interlayer between the GaAs and Fe. For each model, Z-contrast images were simulated with software from Ref. [19] using the relaxed atomic coordinates determined by DFT calculations. Simulations were performed at Scherzer defocus ($\Delta f$ = -40 nm) with collection semiangles of 40 to 175 mrad and projected thicknesses of 4 nm.

The quality of the Fe film, interface, and substrate is revealed by high-resolution TEM (HRTEM) images (Fig. 1) for the spin-LED sample as-grown and after a mild post-growth anneal (200°C for 10 minutes). The as-grown sample showed a SP of 18%, and after heat treatment the SP increased to 26%. In each case, the Fe layer is uniformly thick and well ordered. The phase contrast reveals a 0.565-nm periodicity in the AlGaAs, equivalent to its (001) *d*-spacing. Image simulations of AlGaAs verify the structure and inferred spacing (Fig. 1, inset). The phase contrast of the AlGaAs is periodic from the right to middle part of the image, whereas that of the Fe is periodic from the left to middle. The points at which the phase contrast is no longer periodic in either the Fe or the AlGaAs define the interfacial region between them. The interface is flat and parallel to [110] for both samples. The phase contrast for the 18%-SP sample reveals an interfacial region approximately 0.7-nm thick with some disorder. In comparison, the interfacial region for the 26%-SP sample is approximately 0.5-nm thick with no apparent disorder.

A HAADF image for the 26%-SP sample is shown in Figure 2a. The contrast reveals 0.14-nm periodicities in both Fe and AlGaAs, consistent with their (002) and (004) *d*-spacings, respectively, and indicates that the experimental image resolves atomic columns. The contrast at the interface is best explained by a simple model consisting of one atomic layer of intermixed Fe and As (Fig. 2b, inset). Visual comparison of the experimental image and that for the simulation of the partially intermixed model (Fig. 2b) reveals a compelling match. In addition to the close qualitative correspondence between these experimental and simulated images, we also find good quantitative agreement for intensity profiles both parallel and perpendicular to the interface (Figs. 3 and 4, respectively). We discuss the comparison of two representative profiles in detail below.

Figure 3 shows the intensity profile parallel to the interface, along the line indicated by α to α′ in Fig. 2. The profile consists of alternating high- and low-intensity peaks, corresponding to atoms with relatively high and low atomic number (Fig. 3, solid line). Quantification of the ratios of the peak heights shows that the intensity in the profile follows a $Z^{1.5}$ to $Z^2$ dependency, consistent with Ref. [19], and indicates a row containing alternating Fe and As atoms. This experimental profile is in quantitative agreement with that obtained from the simulation of the partially intermixed model (Fig. 3, dashed line).

A thickness-corrected intensity profile across the interface (indicated by β to β′ in Fig. 2) shows that the columns of Fe, Ga, and As atoms plot as peaks (Fig. 4, red curve), and the relative heights reflect differences in their atomic number. Peaks 6 through 2 have the lowest intensity, consistent with columns of Fe; 0, and -4 contain the highest intensity in the image, indicating columns of As atoms; and peaks -1 and -5 contain intermediate intensity, indicating columns of Ga atoms. Quantification of the ratios of the peak heights shows that the intensity of this profile also follows a $Z^{1.5}$ to $Z^2$ dependency and confirms the identity of the atomic columns. Position 1 indicates no peak intensity between 2 and 0, i.e., atoms do not occupy the space between Fe and As. The broad, low-intensity peak between -2 and -3, is a probe-tailing effect from the electron beam, which we have confirmed by comparison to the simulations. The relative peak intensities and positions are in excellent agreement with those obtained from the partially intermixed interface model, shown as the top inset and curve γ-γ′ in Fig. 4.

In light of a recent report [20] that $Fe_3GaAs$ may form during the growth of Fe on GaAs at temperatures as low as -15ºC, we also considered the possibility of this alloy as an interface phase. We used DFT to relax the structure of a simple abrupt interlayer consisting of five atomic layers of $Fe_3GaAs$ between, and in registry with, the GaAs and Fe. $Fe_3GaAs$ has a hexagonal pseudocubic structure [21] that is well lattice-matched to GaAs and Fe. Moreover, the relaxed atomic positions for this thin interlayer (Fig. 4, bottom inset) appear similar to those of bulk Fe in the [-110] projection (cf., Fig. 4, top inset). We compared the simulated images and intensity profiles of the $Fe_3GaAs$ interlayer with experimental-HAADF images of the annealed sample (26% SP) to test whether it could have formed.

The intensity profile of the simulated $Fe_3GaAs$ interlayer most similar to that of the experimental profile β to β′ is shown as the lower curve (green) in Figure 4 (indicated by δ to δ′ in the ball-and-stick model). A distinguishing feature of the $Fe_3GaAs$ phase is the intensity of the atomic columns that contain As and Ga atoms in addition to Fe (peaks 2 and 4). These intensities are two to four times higher than those of the pure Fe columns (peaks 1, 3, and those ≥ 5). In the experimental profile of the annealed sample (Fig. 4, red curve β to β′), the corresponding columns (positions 2 through 6) show an absolute intensity and variation that are too small to be consistent with anything other than Fe. A second distinguishing feature of the $Fe_3GaAs$ simulation is the intensity at position 1, which corresponds to a column of Fe atoms not present in the partially intermixed interface (Fig. 4, top). The experimental profile shows a minimum at this position, indicating a vacancy, consistent with the simulation of the partially intermixed interface (cf., Fig. 3, blue and red curves). We therefore conclude that $Fe_3GaAs$ does not reflect the true interface structure of the annealed sample. The partially intermixed interface is the more accurate representation.

Based on these detailed comparisons of experimental and simulated images, we conclude that the interface of the annealed sample is ordered and coherent, with intermixing of the Fe and AlGaAs occurring on a single atomic plane (α−α′), resulting in an interface with alternating Fe and As atoms (Fig. 2b). We performed the same experiment for the as-grown 18%-SP sample and found that the intensity of the atom columns is intermediate between Ga and As, indicating chemical disorder. Moreover, the HRTEM image indicates structural disorder that extends over approximately 5 atomic planes (Fig. 1a). Such disorder precludes assigning a specific interface structure to this sample.

We attribute the 44% increase in spin-injection efficiency of the annealed sample to the greater tunneling efficiency for spin-polarized electrons across the chemically and structurally coherent, annealed interface of Fig. 2b. These results are consistent with theoretical work indicating that a reduction in the lattice periodicity at the interface can lead to suppression of spin polarization [22]. First-principles transport calculations have also shown that band *symmetry* plays an important role in spin injection from a metal into a semiconductor [23, 24]. Strong spin filtering is expected to occur at the Fe/GaAs(001) interface, because the $\Delta_1$ symmetry of the bulk Fe majority-spin state near $E_F$ matches that of the bulk GaAs band-edge states, while the symmetry of the Fe minority-spin band does not, resulting in preferential transmission of majority-spin electrons from the Fe. Recent band-structure calculations [25] show this explicitly for the Fe/GaAs(001) abrupt interface model of Ref. [18] – the propagating state of $\Delta_1$ symmetry (i.e., the Fe majority-spin band) decays relatively slowly into the GaAs, promoting transmission, while the states of $\Delta_{2'}$ and $\Delta_5$ symmetry (i.e., the Fe minority-spin bands) decay much more quickly, suppressing transmission of minority spin carriers. We have applied this analysis to the first three interface models reported in Ref. [18]: abrupt, partially intermixed (Fig. 2b), and fully intermixed. Our results show no significant change in the $\Delta_1$ decay rate between the abrupt and partially intermixed, suggesting that both should enable highly polarized spin injection. However, for the fully intermixed model we find a significantly faster decay of the majority spin $\Delta_1$ state into the GaAs, suggesting lower spin polarization of the injected carriers. While this type of calculation cannot address the disorder which probably exists at the interface of the as-grown sample (SP = 18%), these results qualitatively support our experimental observation that spin-injection efficiency is positively correlated with the coherence and abruptness of the interface.

This work was supported in part by the Office of Naval Research, the DARPA SpinS program, and core programs at NRL. TJZ, IŽ, and GK gratefully acknowledge support from the NRL-NRC program. Computations were performed at the DoD Major Shared Resource Center at ASC.

**Figure Captions**

**Figure 1** HRTEM images ([-110] cross section) of an Fe/AlGaAs spin-LED sample. (**a**) As-grown, exhibiting 18% SP, and (**b**) following a mild post-growth anneal, exhibiting a 26% SP. Image simulations are inset (white brackets). The rectangles on the bottom of the images indicate Fe, AlGaAs, and the interfacial region (gray box with cross). Scale bars equal 1.0 nm. The contrast variation across the Fe regions is a result of changes in thickness.

**Figure 2** HAADF images of Fe/AlGaAs [-110]. (**a**) Experimental image from a sample with 26% SP. The arrowheads, $\alpha$ to $\alpha'$ and $\beta$ to $\beta'$, respectively indicate the location and direction of the line profiles shown in Figures 3 and 4. (**b**) Simulated image of a partially intermixed interface [18] and inset ball-and-stick model from which it was calculated (Fe atoms appear in yellow, As in blue, and Ga in red). The scale bar equals 0.5 nm.

**Figure 3** [110] intensity profile parallel to the interface corresponding to positions marked $\alpha$ and $\alpha'$ in Figure 2. The profile from the experimental image is shown as a solid line; that from the simulated image is dashed.

**Figure 4** [001] intensity profiles perpendicular to the interface of the experimental and simulated HAADF images with corresponding model structures. (**Top**) Ball-and-stick model of the partially intermixed interface [18] in the [-110] projection (Fe atoms appear in yellow; As in blue; and Ga in red). The intensity profile of the simulated HAADF

image (sim) from the partially intermixed interface appears in blue (shown as γ to γ′ in its model). The intensity profile from the experimental image (exp) is shown in red and corresponds to the positions marked β and β′ in Figure 2. The intensity profile of the simulated HAADF image (sim) of the $Fe_3GaAs$ interlayer appears in green (shown as δ to δ′ in its model). (**Bottom**) Ball-and-stick model of the $Fe_3GaAs$ interlayer in the [-110] projection (color scheme is the same as top). Annotations are discussed in the text. Minor differences between experiment and simulation include: the deep minimum in the simulation at position 1 due to the greater resolution; the small peak at position 1 due to probe tails; and slight offset of peaks at position 2.

**Figure 1**

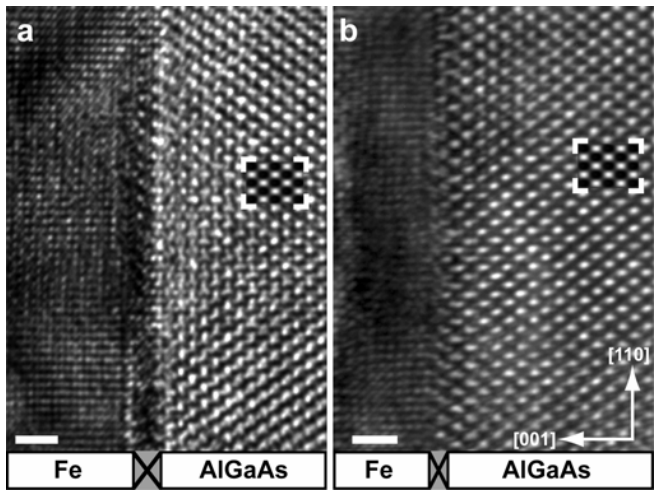

**Figure 2**

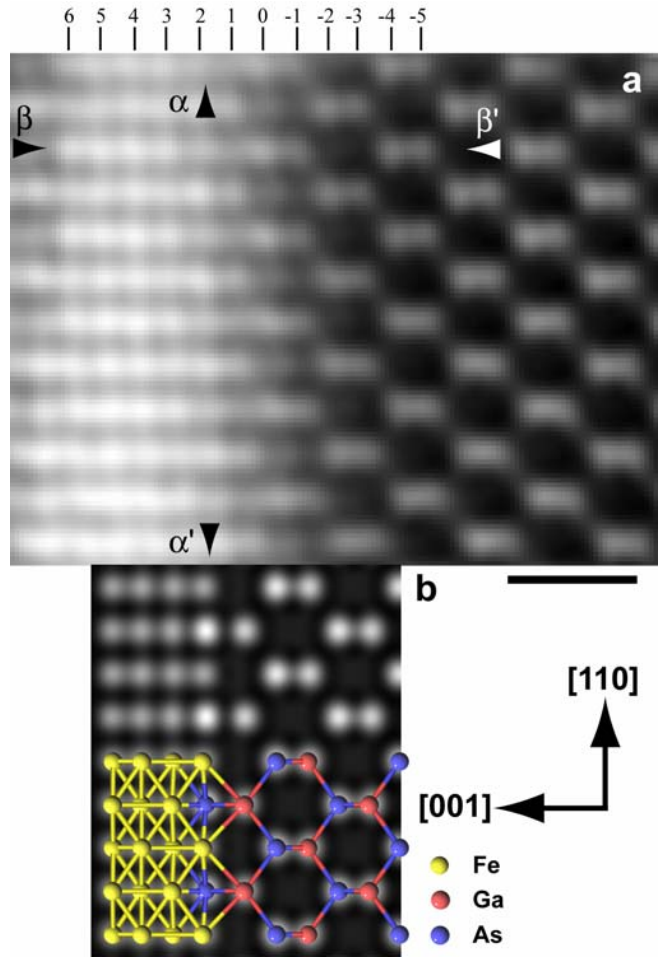

**Figure 3**

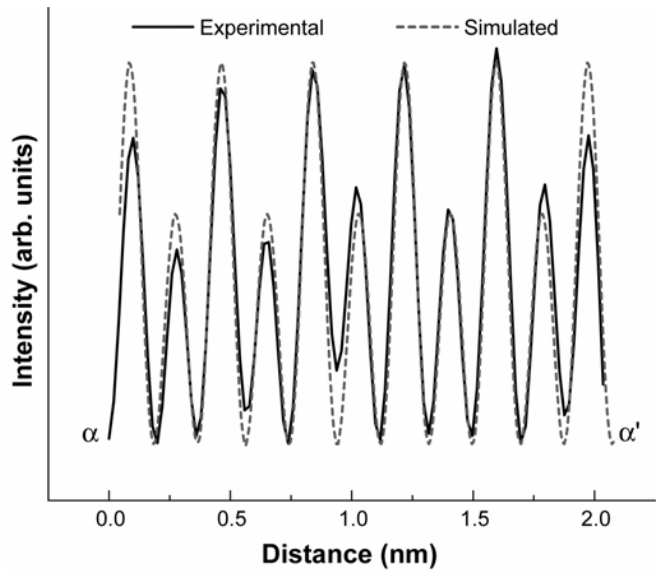

**Figure 4**

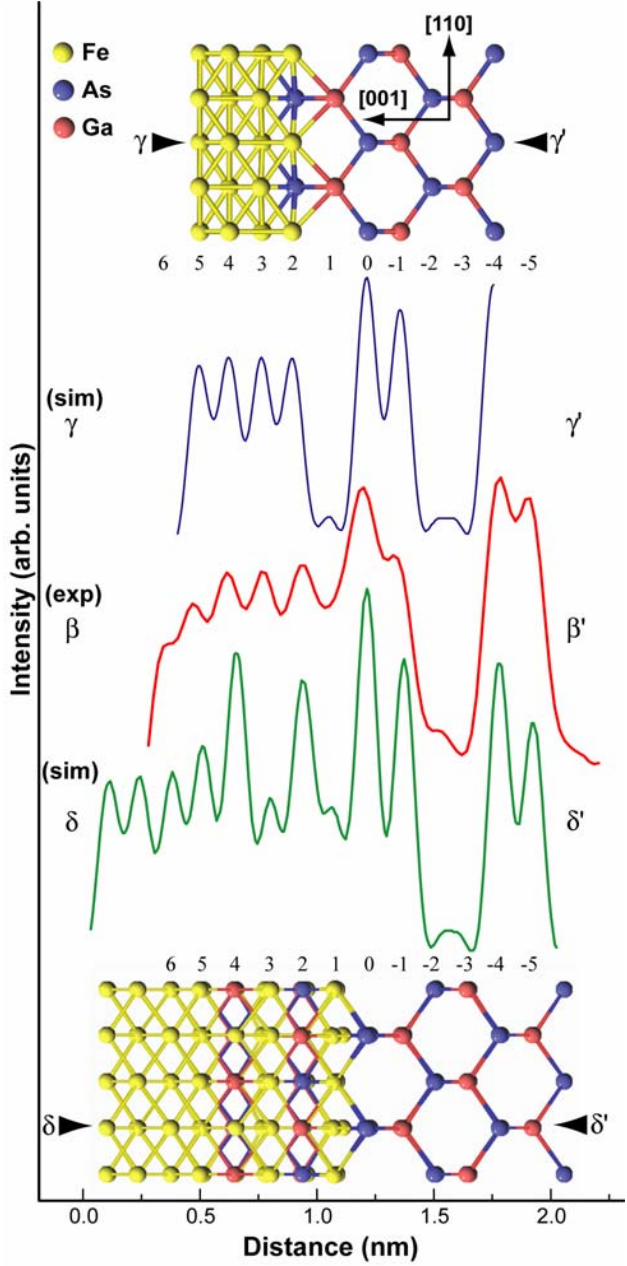